\documentclass{revtex4-2} 

\usepackage{amsmath}  
\usepackage{amsfonts} 
\usepackage{graphicx} 
\usepackage{esvect}
\usepackage{mathtools}
\usepackage{subcaption}
\usepackage{xcolor}
\usepackage{appendix}

\usepackage{siunitx} 
\usepackage{placeins}

\usepackage{float}

\usepackage{pbox}

\usepackage{url}

\usepackage{xcolor}
\pagecolor{white}

\usepackage{mathtools}
\makeatletter
\newcommand{\vast}{\bBigg@{4}}
\newcommand{\Vast}{\bBigg@{5}}
\makeatother

\begin{document}


\title{On the  Electrostatic Interaction \\ between Point Charges due to Dielectrical Shielding}

\author{Long T. Nguyen}
\affiliation{High School for Gifted Students, Hanoi University of Science, Vietnam National University, 182 Luong The Vinh Str., Thanh Xuan, Hanoi 100000, Vietnam.}

\author{Tuan K. Do}
\affiliation{Department of Mathematics, Princeton University, Princeton, NJ 08544, USA.}

\author{Duy V. Nguyen} 
\affiliation{Phenikaa Institute for Advanced Study, Phenikaa University, Hanoi 100000, Vietnam.}

\author{Trung V. Phan}
\email{Correspoding author: trung.phan@yale.edu}
\affiliation{Department of Molecular, Cellular, and Developmental Biology, \\ Yale University, New Haven, CT 06520, USA.}

\date{\today}

\begin{abstract}
How will the electrostatic interaction between two point charges change if they are shielded from the other by a dielectrical slab? While the physical setting of this electromagnetic problem is relatively simple, it is easy to be wronged and the correct solution is surprisingly complicated. Here we will show a general answer using the method of images, in which the electrical field are not found by solving the Poisson's equation but by superposing an infinite number of image charges to recurrently satisfy all interfaces' boundary conditions. We also obtain analytical and algebraic results in some special cases. 
\end{abstract}

\maketitle

\section{The Common Mistake using Naive Spatial Expansion}

\begin{figure}[!htb]
\centering
\includegraphics[width=0.60\textwidth]{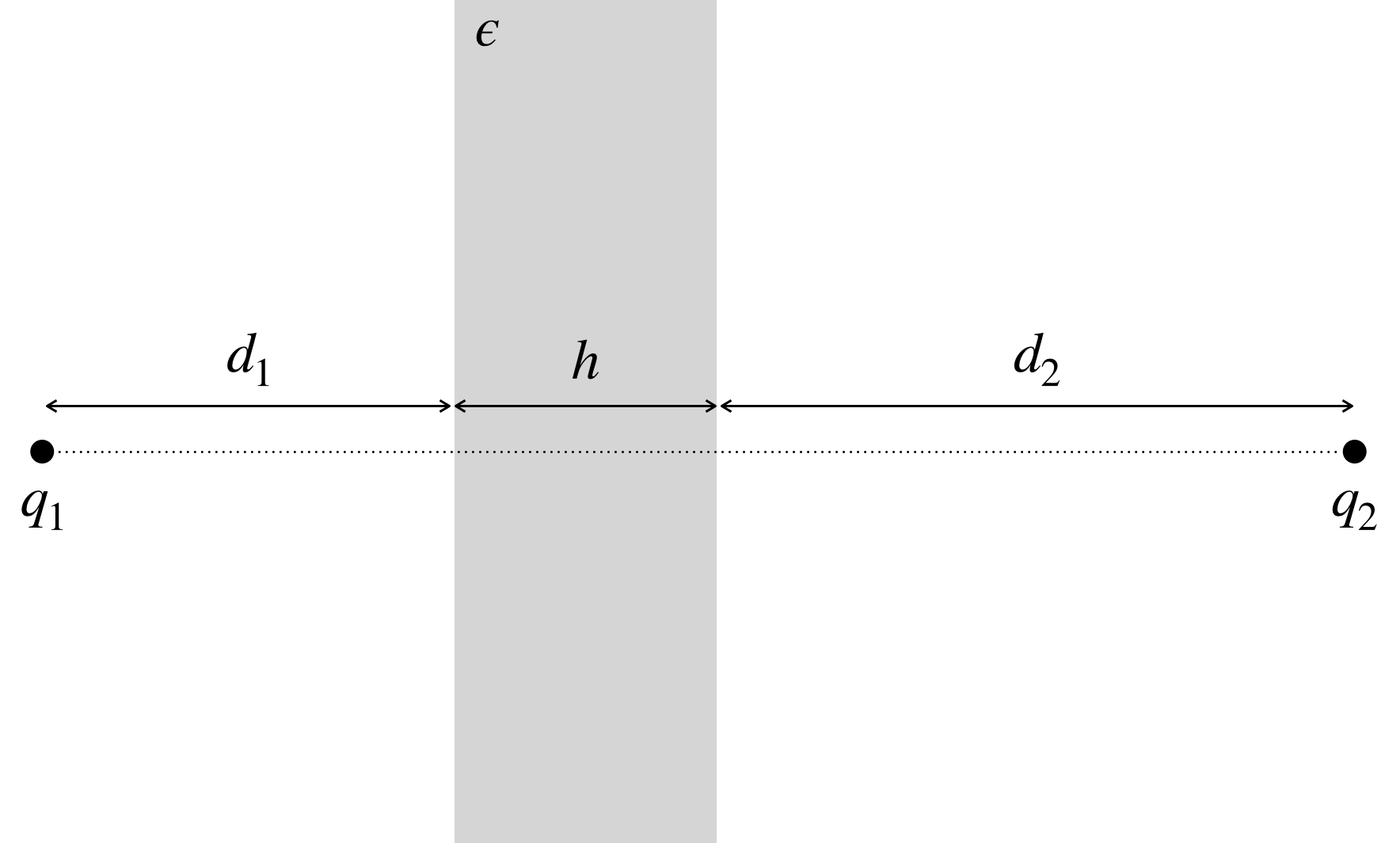}
\caption{An infinitely large dielectric slab is inserted in between two point charges.} 
\label{fig01}
\end{figure}

We are interested in determining the electrostatic forces that act on the two point charges $q_1$ and $q_2$ placed in vacuum in the presence of an infinitely large dielectric slab of thickness $h$ and relative dielectric constant $\epsilon$ inserted in between, as shown in Fig. \ref{fig01}. The distance between the charges and the slab are $d_1$ and $d_2$. While this is a simple setting that represents the phenomena of electromagnetic shielding, which has a wide variety of applications \cite{celozzi2008electromagnetic}, it is curious that the answer for this question cannot be found in  textbooks (and also hard to find in the literature, e.g. as a Green's functional integration \cite{barcellona2018manipulating}). 

Let us briefly mention a common mistake. In many undergraduate introductory and advance high school physics courses \cite{purcell2013electricity,landau2013electrodynamics,stratton2007electromagnetic}, the students are taught that the electrostatic force $f$ between charges inside a medium of relative dielectric constant $\epsilon$ is given by the Coulomb's equation:
\begin{equation}
f = \frac{1}{4\pi \epsilon_0 \epsilon} \frac{q_1 q_2}{d^2} \ , 
\label{coulomb_equation}
\end{equation}
where $\epsilon_0$ is the dielectric constant of vacuum, $q_1$ and $q_2$ are two point charges of interests and $d$ is the distance between them. These electrostatic forces acting on the two charges are opposite in direction but equal in magnitude, and this value is the same as if the charges are distance $\sqrt{\epsilon}d$ apart in vacuum. Due to that, when dealing with the setting in Fig. \ref{fig01}, many has incorrectly assumed that the presence of the dielectric slab is no different than having the region of space it is filling expands by the factor $\sqrt{\epsilon}$, thus arrive at:
\begin{equation}
f = \frac{1}{4\pi \epsilon_0} \frac{q_1 q_2}{\left( d_1 + \sqrt{\epsilon}h + d_2 \right)^2}   \ .
\label{wrong_equation}
\end{equation}
Here, we will show that not only is this value wrong, but also these forces are non-reciprocal: the magnitudes of the forces acting on charge $q_1$ and $q_2$ are in general not the same, $|f_1| \neq |f_2|$. The image charges configuration and the calculation is quite non-trivial, and to the best of our knowledge it has not been carried out and analyzed in much details. This paper is set out to fill this gap, in 
a geometrical way -- using the method of images \cite{hammond1960electric, jackson1999classical}) -- that highlights how the boundary conditions for the electrical field at the dielectric interfaces can be satisfied, recurrently. For other applications of this in different electrostatic settings, see e.g. \cite{lindell1992electrostatic,lindell1992electrostaticL,sten1992electrostatic,nikoshkinen1995image, lindell2001electrostatic}.

\section{Infinite Image Charges}

The key understanding for this problem is to realize that the dielectric slab doesn't effectively expand space. Physically, what happens is that the charges inside the dielectric slab can redistribute and create surface charge densities on the interfaces which shield the external electrical field.

\begin{figure*}[!htb]
\centering
\includegraphics[width=\textwidth]{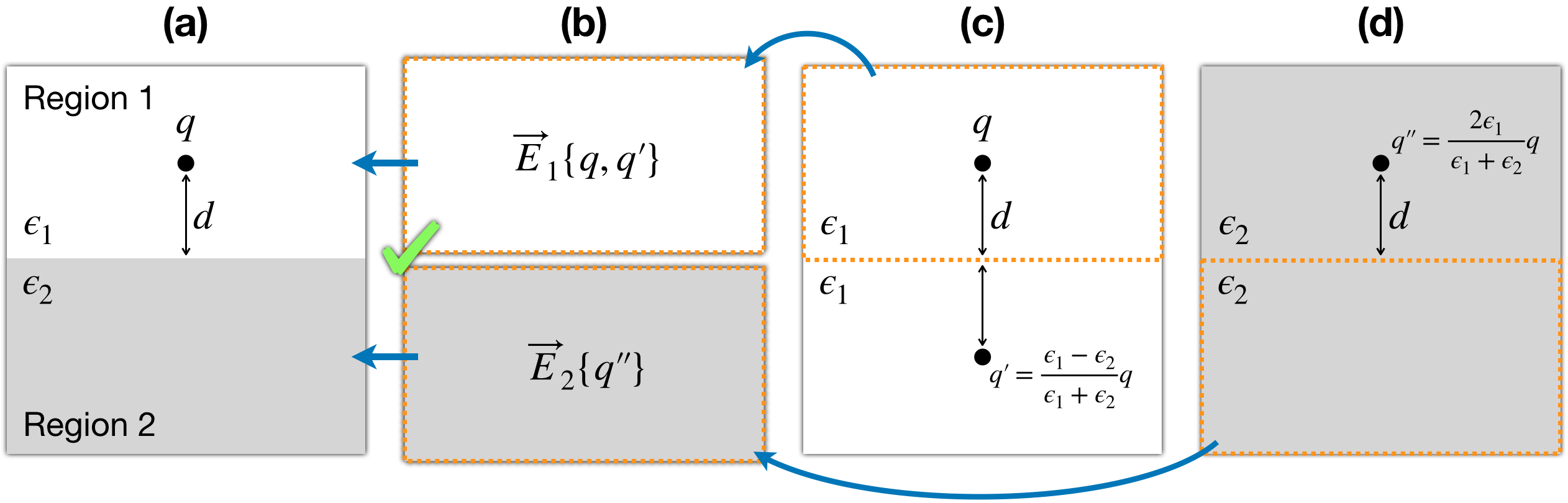}
\caption{(a) The space is divided into two regions, Region 1 and Region 2, having different dielectric constants by an infinite-planar interface. A point charge $q$ is introduced inside Region 1. (b) The stitching procedure for electrical field, gluing $\vec{E}_1\{q,q'\}$ in Region 1 and $\vec{E}_2 \{q''\}$ in Region 2 so that all boundary conditions at the interface can be satisfied. (c) The electrical field $\vec{E}_1\{q,q'\}$ in Region 1. (d) The electrical field $\vec{E}_2\{q''\}$ in Region 2.}
\label{fig02}
\end{figure*}

First, let us recall the standard method of images for a planar interface between two dielectric mediums. Consider the space is filled with mediums of different relative dielectric constants: $\epsilon_1$ for Region 1 and $\epsilon_2$ for Region 2, separated by an infinite-planar interface (see Fig. \ref{fig02}a). Place a point charge $q$ in Region 1 at distance $d$ away from the interface. The electrical field can then be determined by a stitching procedure, in which we say {\it the object $q$ through the interface gives two images $q$ and $q'$} (see Fig. \ref{fig02}b): the field $\vec{E}_1\{q,q'\}$ in Region 1 which is created by the original point charges $q$ and an image charge $q'=(\epsilon_1 - \epsilon_2)/(\epsilon_1+\epsilon_2) \times q$ located at the mirror-reflected position of charge $q$ through the interface (see Fig. \ref{fig02}c), and the field $\vec{E}_2\{q''\}$ in Region 2 which is created by an image charge $q''=2 \epsilon_1/(\epsilon_1+\epsilon_2)\times q$ located at the same position with $q$ (see Fig. \ref{fig02}d). We can check by direct substitution that not only does this stitched electrical field satisfy the Poisson's equation everywhere but is also consistent with the boundary conditions at the interface: for the normal component $\epsilon_1 E_{\parallel 1} = \epsilon_2 E_{\parallel 2} $ and for the tangential component $E_{\perp 1} = E_{\perp 2} $. We say {\it $\vec{E}_1\{q,q'\}$ and $\vec{E}_2\{q''\}$ match through the interface}.

\begin{figure*}[!htb]
\centering
\includegraphics[width=\textwidth]{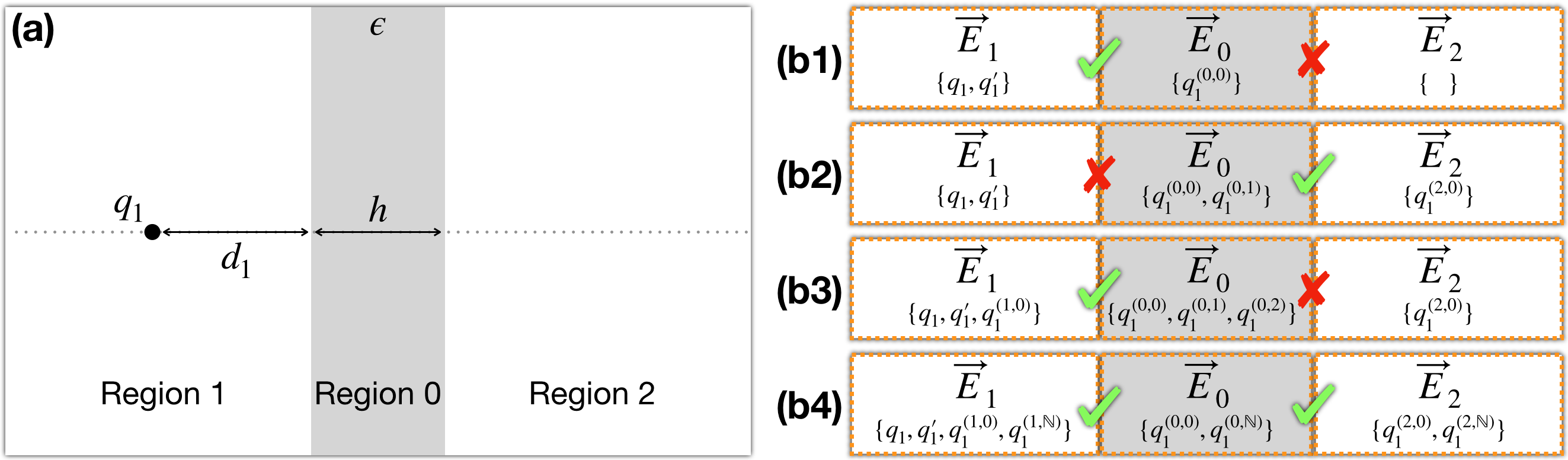}
\caption{(Step 0)} 
\label{fig03}
\end{figure*}

Secondly, consider a system placed in vacuum consists of a point charge $q_1$ at distance $d_1$ from an infinitely large dielectric slab of thickness $h$ and relative dielectric constant $\epsilon$. The setting here is similar to \cite{sometani2000image}. The slab divides space into three regions: Region $1$ has the charge $q_1$, Region $0$ is the inside of the slab and Region $2$ is the rest of the space (see Fig. \ref{fig03}). There are now two interfaces which boundary conditions need to be satisfied. To find the electrical field everywhere in space, we can use the method of images as mentioned above, recurrently for an infinite number of times, as followed:

\begin{itemize}
\item {\bf Step 1:} The object $q_1$ in Region 1, through the interface between Region 1 and Region 0, gives two images: $q_1'$ influences on Region 1 and $q_1^{(0,0)}$ influences on Region 0. While $\vec{E}_1\{q_1,q_1'\}$ and $\vec{E}_0\{q_1^{(0,0)}\}$ match through the interface between Region 1 and Region 0, $\vec{E}_0\{q_1,q_1'\}$ and $\vec{E}_2\{ \ \ \}$ do not match through the interface between Region 0 and Region 2 due to $q_1^{(0,0)}$ influences in Region 0 (see Fig. \ref{fig03}b1).

\item {\bf Step 2:} The object $q_1^{(0,0)}$ influences Region 0, through the interface between Region 0 and Region 2, gives two images: $q_1^{(0,1)}$ influences on Region 0 and $q_1^{(2,0)}$ influences on Region 2. While $\vec{E}_0\{q_1^{(0,0)},q_1^{(0,1)}\}$ and $\vec{E}_2\{q_1^{(2,0)}\}$ match through the interface between Region 0 and Region 2, $\vec{E}_0\{q_1^{(0,0)},q_1^{(0,1)}\}$ and $\vec{E}_1\{q_1,q_1'\}$ do not match through the interface between Region 0 and Region 1 due to $q_1^{(0,1)}$ influences in Region 0 (see Fig. \ref{fig03}b2).

\item {\bf Step 3:} The object $q_1^{(0,1)}$ influences Region 0, through the interface between Region 0 and Region 1, gives two images: $q_1^{(0,2)}$ influences on Region 0 and $q_1^{(1,0)}$ influences on Region 1. While $\vec{E}_0\{q_1^{(0,0)},q_1^{(0,1)},q_1^{(0,2)}\}$ and $\vec{E}_1\{q_1,q_1',q_1^{(1,0)}\}$ match through the interface between Region 0 and Region 1, $\vec{E}_0\{q_1^{(0,0)},q_1^{(0,1)},q_1^{(0,2)}\}$ and $\vec{E}_2\{q_1^{(2,0)}\}$ do not match through the interface between Region 0 and Region 2 due to $q_1^{(0,2)}$ influences in Region 0 (see Fig. \ref{fig03}b3).

\item {\bf Step 4 forward:} Note that the end of Step 3 is similar to the end of Step 1, with $q_1^{(2,0)}$ instead of $q_1^{(0,0)}$. Thus we can keep repeating the steps infinitely many times, and get the electrical field converges into $\vec{E}_1\{ q_1, q_1',q_1^{(1,0)},q_1^{(1,\mathbb{N})}\}$, $\vec{E}_0\{ q_1^{(0,0)},q_1^{(0,\mathbb{N})}\}$, $\vec{E}_2\{q_1^{(2,0)},q_1^{(2,\mathbb{N})}\}$ where $\mathbb{N}=\{ 1,2,3, ...\}$, which now should match on both interfaces since the mismatch gets smaller and smaller after each step (see Fig. \ref{fig03}b4).
\end{itemize}

\begin{figure*}[!htb]
\centering
\includegraphics[width=\textwidth]{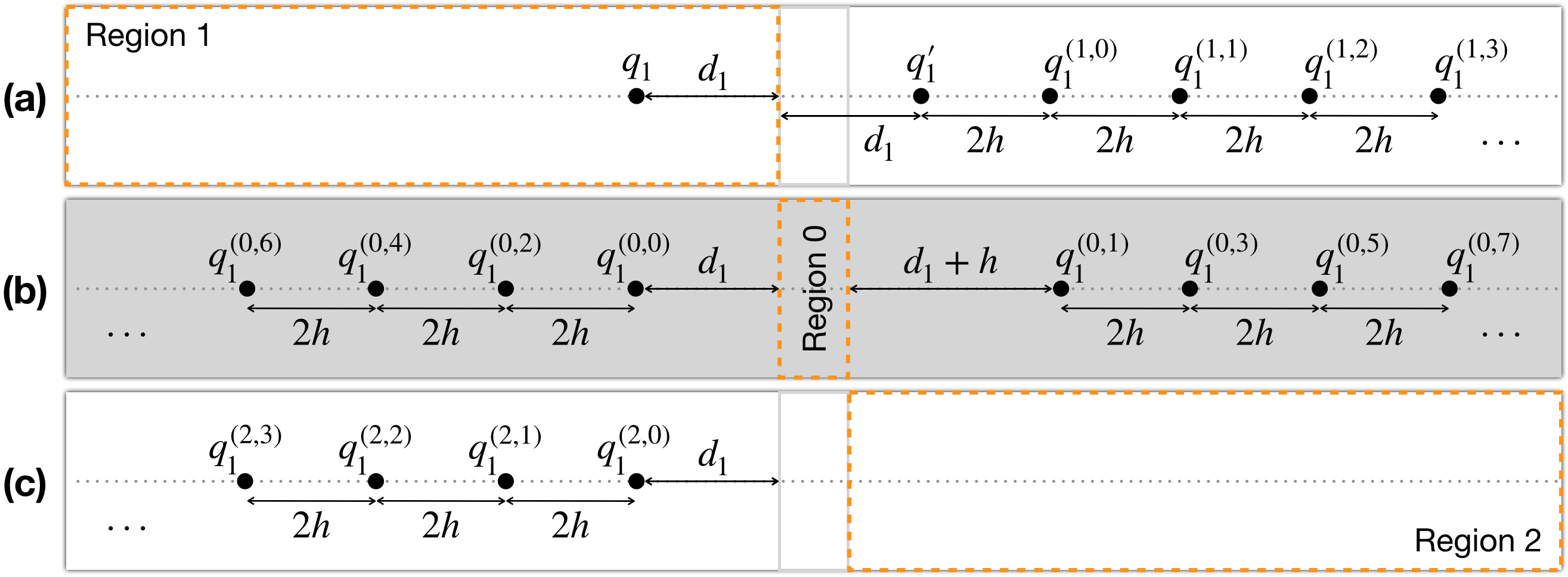}
\caption{(a) $\vec{E}_1\{ q_1, q_1',q_1^{(1,0)},q_1^{(1,\mathbb{N})}\}$. (b) $\vec{E}_0\{ q_1^{(0,0)},q_1^{(0,\mathbb{N})}\}$. (c) $\vec{E}_2\{q_1^{(2,0)},q_1^{(2,\mathbb{N})}\}$.}
\label{fig04}
\end{figure*}

In more details, the electrical field $\vec{E}_1$ in Region 1 is as if the whole space has an uniform dielectric constant $1$ and there is an infinite series of point charges: the original charge $q_1$, an image charge $q_1'$ placed outside Region 1 at distance $2d_1$ away from $q_1$ and image charges $q_1^{(1,k)}$ where $k=0,1,2,3,...$ placed outside Region 1 at distance $2(d_1+h) + 2kh$ away from $q_1$ (see Fig. \ref{fig04}a):
\begin{equation}
q_1' = -\frac{\epsilon-1}{\epsilon+1} q_1 \ , \ q_1^{(1,k)} = \epsilon \left( \frac{2}{\epsilon+1} \right)^2 \left( \frac{\epsilon-1}{\epsilon+1} \right)^{2k+1} q_1 \ .
\label{region1_charge1}
\end{equation}
The electrical field $\vec{E}_0$ in Region 0 is as if the whole space has an uniform dielectric constant $\epsilon$ and there is an infinite series of point charges: image charges $q_1^{(0,k)}$ where $k=0,2,4,...$ placed inside Region 1 at distance $kh$ away from $q_1$ and image charges $q_1^{(0,k)}$ where $k=1,3,5,...$ placed inside Region 2 at distance $2(d_1+h)+(k-1)h$ away from $q_1$ (see Fig. \ref{fig04}b):
\begin{equation}
q_1^{(0,k)} = \left( \frac{2}{\epsilon+1} \right) \left( \frac{\epsilon-1}{\epsilon+1} \right)^{k} q_1 \ .
\label{region0_charge1}
\end{equation}
The electrical field $\vec{E}_2$ in Region 2 is as if the whole space has an uniform dielectric constant $1$ and there is an infinite series of point charges: image charges $q_1^{(0,k)}$ where $k=0,1,2,3,...$ placed inside Region 1 at distance $2kh$ away from $q_1$ (see Fig. \ref{fig04}c):
\begin{equation}
q_1^{(2,k)} = \epsilon \left( \frac{2}{\epsilon+1} \right)^2 \left( \frac{\epsilon-1}{\epsilon+1} \right)^{2k} q_1 \ .
\label{region2_charge1}
\end{equation}
The above geometrical construction for an infinite series of image charges is possible for any (positive) value of $d_1$, $d_2$ and $h$.

\section{The Electrostatic Forces}

\subsection{The general case}

\begin{figure*}[!htb]
\centering
\includegraphics[width=\textwidth]{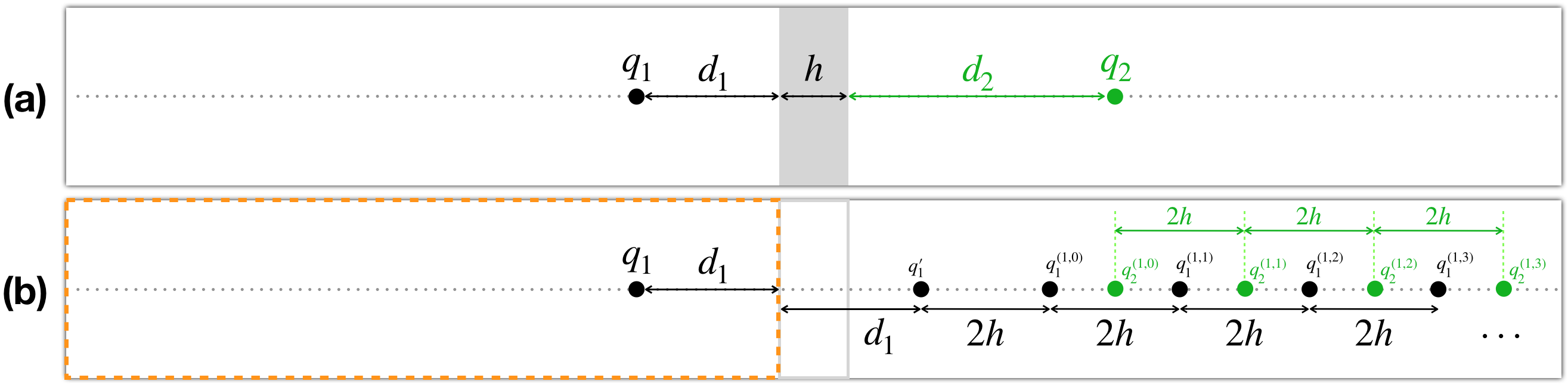}
\caption{(a) Introduce charge $q_2$ on the opposite side of charge $q_1$ with respect to the slab. (b) Superposing the image charges from charge $q_1$ and $q_2$, to get the electrical field in Region $1$.} 
\label{fig05}
\end{figure*}

Finally, now we have enough ingredient to solve the original problem (see Fig. \ref{fig01}), by considering the superposition of all the charges and image charges as we introduced charge $q_2$ at region $2$ distance $d_2$ away from the dielectric slab, on the other side (see Fig. \ref{fig05}a). For example, the electrical field in Region 1 is as if not only are there charge $q_1$ and image charges $q_1'$ $q_1^{(1,k)}$ where $k=0,1,2,...$, but also there are $q_2^{(1,k)}$ placed inside region $2$ at distance $(d_1+h+d_2)+2kh$ away from charge $q_1$ (see Fig. \ref{fig05}b):
\begin{equation}
q_2^{(1,k)} = \epsilon \left( \frac{2}{\epsilon+1} \right)^2 \left( \frac{\epsilon-1}{\epsilon+1} \right)^{2k} q_2 \ .
\label{region1_charge2}
\end{equation}
In other words, $\vec{E}_1 = \vec{E}_1 \{ q_1, q_1',q_1^{(1,0)},q_1^{(1,\mathbb{N})}\} + \vec{E}_1 \{ q_2^{(1,0)},q_2^{(1,\mathbb{N})}\}$. For a sanity check, we calculate the electrical field at the interfaces and show that the boundary conditions are satisfied everywhere on those planes in Supplementary Material Section 1.

Thus the electrostatic force $f_1$ acting on charge $q_1$ can be calculated as an infinite summation series:
\begin{equation}
f_1 = \frac{q_1}{4\pi \epsilon_0} \Bigg( \frac{q_1'}{(2d_1)^2} + \sum^{\infty}_{k=0}  \frac{q_1^{(1,k)}}{\big( 2(d_1+h) + 2kh \big)^2}
+ \sum^{\infty}_{k=0} \frac{q_2^{(1,k)}}{\big( (d_1+h+d_2) + 2kh \big)^2}  \Bigg) \ .
\label{general_force}
\end{equation}
Define the ratios $\alpha_1 = d_1/h$, $\alpha_2=d_2/h$ and define the summation $S_{a}(z)$:
\begin{equation}
S_a(z) = \sum^\infty_{k=0} \frac{z^{2k}}{(a + 2k)^2} \ , 
\label{sum_blob}
\end{equation}
we can rewrite the force equation as:
\begin{equation}
\begin{split}
f_1 &= \frac{q_1}{4\pi \epsilon_0 h^2} \Bigg[ - q_1 \left( \frac{\epsilon - 1}{\epsilon + 1} \right) \frac{1}{ (2 \alpha_1)^2} \\
&+ q_1 \epsilon \left( \frac{2}{\epsilon + 1} \right)^2 \left( \frac{\epsilon - 1}{\epsilon + 1} \right) S_{2(\alpha_1+1)} \left( \frac{\epsilon - 1}{\epsilon + 1} \right)
+ q_2 \epsilon \left( \frac{2}{\epsilon + 1} \right)^2 S_{\alpha_1+1+\alpha_2 } \left( \frac{\epsilon - 1}{\epsilon + 1} \right) \Bigg] \ .
\end{split}
\label{general_sum}
\end{equation}
Similarly, we get $f_2$ acting on charge $q_2$, which magnitude can be shown to be different from that of $f_1$ in general. This means the interaction between these two charges mediated by the dielectric slab is non-reciprocal. At first glance, this seems like a violation of Newton's third law, however it should be noted that not only are interactions between the point-charges but also there are interactions between the point-charges and the dielectric slab. A detail calculation, given in Supplementary Material Section 2, show that Newton's third law still holds.

\subsection{Some special cases}

There are special cases where Eq. \eqref{general_sum} can be represented by special functions and even have a nice short algebraic form. 

The dilogarithm function \cite{cartier2007frontiers} $\text{Li}_2(z) = \sum^{\infty}_{k=1} z^k/k^2$ can be used to rewrite the infinite sum in Eq. \eqref{sum_blob} with finite terms. For examples, when $a \in 2\mathbb{Z}^+$:
\begin{equation}
S_a(z) = \frac{\text{Li}_2 (z^2 ) - \sum^{(a-2)/2}_{k=1} \frac{z^{2k}}{k^2}}{4z^{a}}  \ ,
\end{equation}
and when $a \in 2\mathbb{Z}^+ + 1$:
\begin{equation}
S_a(z) = \frac{\text{Li}_2 (z) - \text{Li}_2 (z^2)/4 - \sum^{(a-3)/2}_{k=0} \frac{z^{2k+1}}{(2k+1)^2}}{z^{a}} \ .
\end{equation}
When $d_1=h=d_2$, we get the ratios $\alpha_1=\alpha_2=1$ and the electrostatic force in Eq. \eqref{general_sum} becomes:
\begin{equation}
\begin{split}
&f_1 = \frac{q_1^2}{4\pi \epsilon_0 h^2} \left( - \frac{\zeta}4 + \frac{(1-\zeta^2) \left( \text{Li}_2(\zeta^2) - \zeta^2 \right) }{4\zeta^3} \right)
\\
& + \frac{q_1 q_2}{4\pi \epsilon_0 h^2} \left( \frac{(1-\zeta^2) \left( 4\text{Li}_2(\zeta) - \text{Li}_2(\zeta^2) - 4\zeta \right) }{4\zeta^3} \right) \ ,
\end{split}
\label{special_d1=h=d2}
\end{equation}
where $\zeta = (\epsilon-1)/(\epsilon+1)$. Consider further simplification with $q_1=q_2=q$, the interaction is now reciprocal $f_1=f_2=f$ and we also arrive at:
\begin{equation} 
f = \frac{q^2}{4\pi \epsilon_0 h^2} \left(1 - \frac1{4\zeta} - \frac1{\zeta^2} - \frac{\text{Li}_2(\zeta)}{\zeta} + \frac{\text{Li}_2(\zeta)}{\zeta^3} \right) \ . 
\label{special_case01_details}
\end{equation}
Note that this setting is highly-symmetric, which makes directly solving the Poisson's equation much easier than in a general case. In Supplementary Material Section 3, we derive Eq. \eqref{special_case01_details} by dealing with the partial differential equation, thus show the consistency between the two different approaches (the geometrical method described in this paper and the calculus method which might be more familiar to most students of physics).

For the relative dielectric constant $\epsilon=3$, we get $\zeta=1/2$. Since $\text{Li}_2(1/2)=\pi^2/12 - \ln^22/2$, \cite{loxton1984special} we can arrive at the following algebraic form:
\begin{equation} 
\begin{split}
f = \frac{q^2}{4\pi \epsilon_0 h^2} \times \frac{\pi^2-7 - 6 \ln^2 2 }2 \approx \frac{q^2}{4\pi \epsilon_0 h^2} \times \left( -6.6 \times 10^{-3} \right) \ . 
\end{split}
\end{equation}
The ``$-$'' sign indicates that the electrostatic forces are pulling the charges toward the dielectric slab. Compare with Eq. \eqref{wrong_equation} which comes from the naive assumption of expanding space:
\begin{equation} 
f = \frac{q^2}{4\pi \epsilon_0 (2+\sqrt{3})^2 h^2} \approx \frac{q^2}{4\pi \epsilon_0 h^2} \times \left( +7.2 \times 10^{-2} \right) \ , 
\label{epsilon_3}
\end{equation}
this wrong answer has incorrect sign and the interaction strength is off by an order of magnitude. 

There are other interesting limits that can be read-off easily from Eq. \eqref{special_case01_details}. For the relative dielectric constant satisfies $|\epsilon - 1| \ll 1$, we have $\zeta \approx (\epsilon-1)/2$, $\text{Li}_2(\zeta) \approx (\epsilon-1)/2$  and get the approximation:
\begin{equation} 
f  \approx \frac{q^2}{4\pi \epsilon_0 h^2} \left(\frac19 - \frac{3(\epsilon-1)}{32} \right) > 0 \ ,
\label{epsilon_1}
\end{equation}
which is consistent with the slab vanishing as $\epsilon \rightarrow 1$. For the relative dielectric constant satisfies $\epsilon \gg 1$, we have $\zeta \approx 1$, $\text{Li}_2(\zeta) \approx \pi^2/6$ and get the approximation:
\begin{equation} 
f  \approx \frac{q^2}{4\pi \epsilon_0 h^2} \left( -\frac14 \right) = -\frac{q^2}{4\pi \epsilon_0 (2d_1)^2} < 0 \ ,
\label{epsilon_gg_1}
\end{equation}
which is consistent with the slab being a conductor (the interaction is mostly between $q_1$ and its image charges $q_1'\approx -q_1$ while the electrical field from $q_2$ is perfectly shielded out).

\begin{figure}[!htb]
\centering
\includegraphics[width=0.80\textwidth]{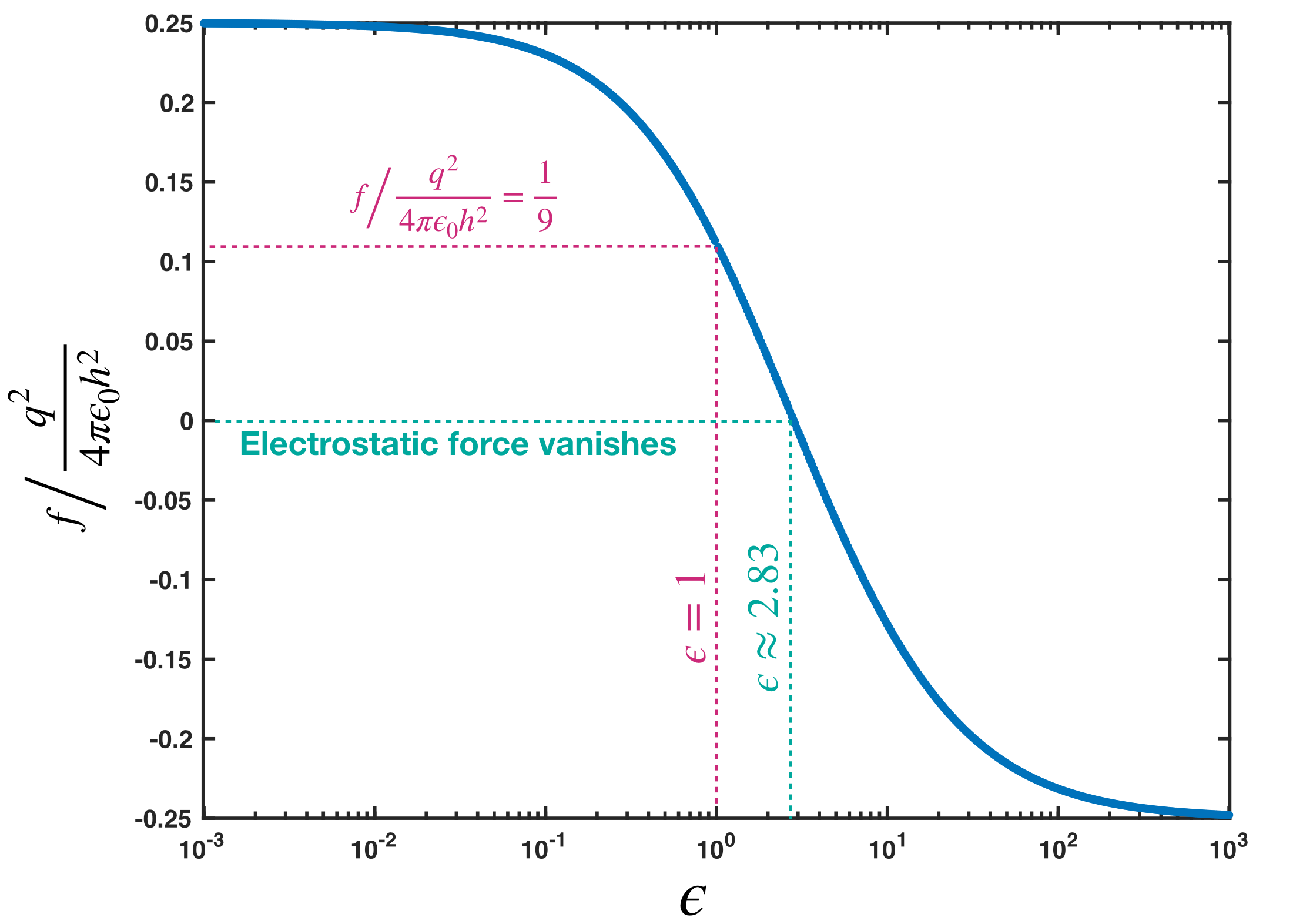}
\caption{The electrostatic force $f$ as a function of dielectric constant $\epsilon$, using Eq. \eqref{special_case01_details}.} 
\label{fig06}
\end{figure}

There exists a value $\epsilon$ in which the electrostatic forces acting on the charges vanish, which corresponds to the perfect shielding scenario. From \eqref{epsilon_3}, \eqref{epsilon_1} and \eqref{epsilon_gg_1}, that value of $\epsilon$ can be guessed to be close to $\epsilon=3$. Try the ansatz $\epsilon = 3+\delta$ in which $\delta \ll 3$, then in the leading order of $\delta$ Eq. \eqref{special_case01_details} can be written as:
\begin{equation}
f \approx \frac{q^2}{4\pi \epsilon_0 h^2} \times \Bigg( \frac{\pi^2-7 - 6 \ln^2 2  }2 + \frac{-11\pi^2 + 51 + 66 \ln^2 2 + 36 \ln 2}{24}\delta \Bigg) \ ,
\end{equation}
thus $f = 0$ can be satisfied around:
\begin{equation}
\delta = \frac{12\pi^2 - 84 - 72 \ln^2 2}{-11\pi^2 + 51 + 66 \ln^2 2 + 36 \ln 2} \approx -0.1744 ,
\end{equation}
which results in $\epsilon \approx 2.826$. A numerical investigation of Eq. \eqref{special_case01_details}, which gives Fig. \ref{fig06}, is indeed in good agreement as $f=0$ when $\epsilon \approx 2.831$.

\section{Conclusion}

We have calculated the interaction between charges after separating them by a dielectric shielding, using the method of images (which should be familiar to most undergraduate students majored in physics and electrical engineering) and representing the answer by an infinite summation series (which can be numerically evaluated). We hope that this paper can fill in the gap often left in introductory physics course or at the very least eliminate the incorrect understanding on a simple setting of great relevant to many industrial applications. Last but not least, we would like to mention the recent 2020 work by Bossa et al \cite{bossa2020integral} which use a different setting but somewhat in a similar spirit (calculating the interaction between two point-charges both inside the same a lipid layer, which assumed to be dielectric) and has a concrete biological implication.

\section{Acknowledgement}

We thank Tung X. Tran, Nam H. Nguyen and the xPhO journal club for their support to share this finding to a wider audience. We also thank Hoi K. Pham for comments and initial interests.

\appendix

\section{Checking the Boundary Conditions}

\begin{figure}[!htb]
    \centering
    \includegraphics[width=0.5\textwidth]{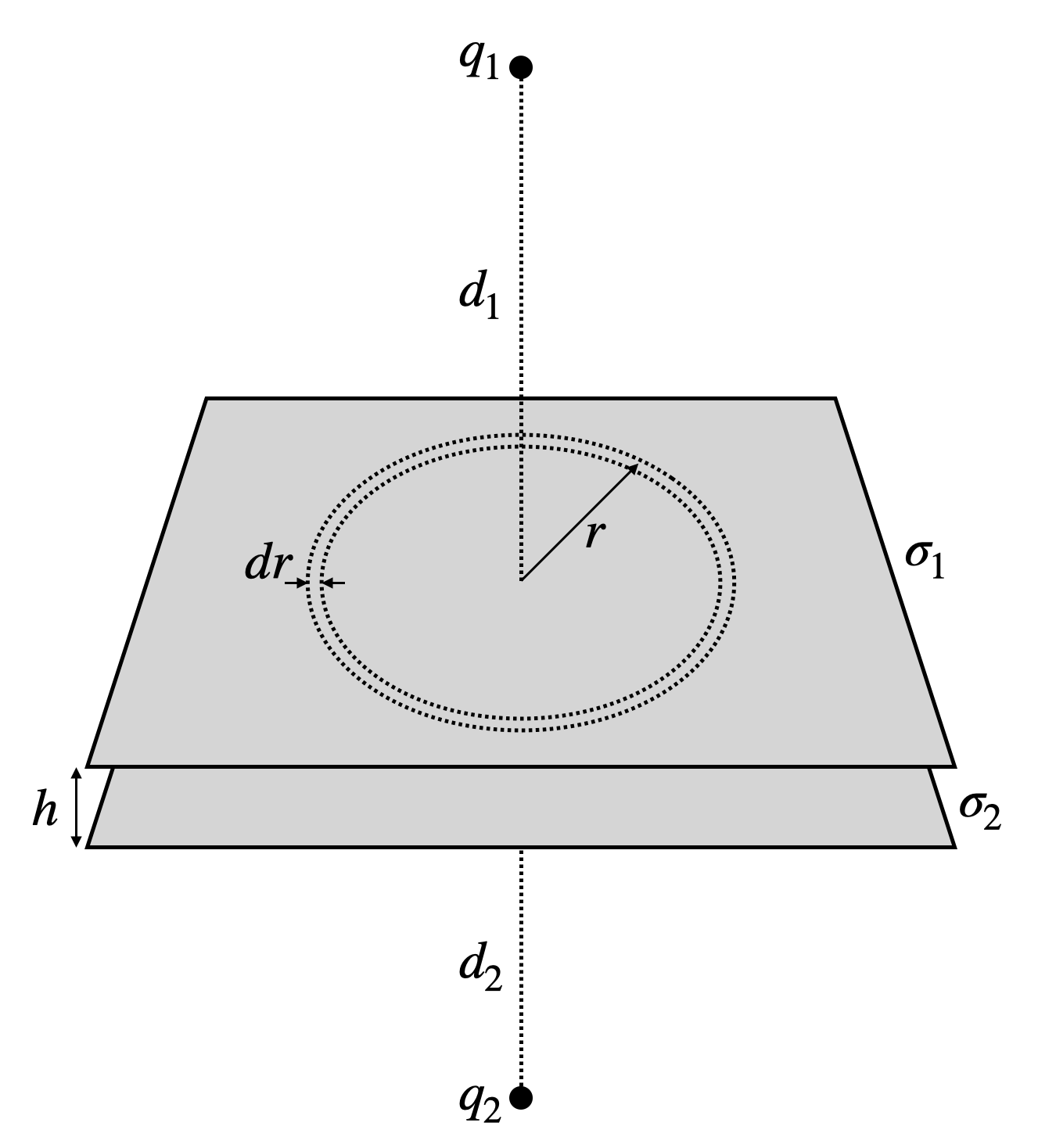}
    \caption{The cylindrical coordinate system we use in Section 1 and Section 2.}
    \label{figA5}
\end{figure}

We only have to check the boundary conditions at one of the interfaces, because for the other it is similar. Let us pick the interface between Region 1 and Region 0, and choose a cylindrical coordinate system in which the central axis passes through both point-charges $q_1$ and $q_2$, $r$ is the radial distance away from that axis (see Fig. \ref{figA5}).

\subsection{Tangential Condition: $E_{\parallel 1} = E_{\parallel 0}$}

\begin{itemize}
    \item On the interface, inside Region 1: \\
    \begin{equation}
        \begin{split}
            E_{\parallel 1} &= \frac{1}{4\pi\epsilon_0} \Bigg\{ \frac{q_1 r}{[d_1^2+r^2]^{3/2}} + \frac{q_1' r}{[d_1^2+r^2]^{3/2}} + \sum_{k=0}^\infty  \frac{q_1^{(1,k)} r}{[(d_1+2h+2kh)^2+r^2]^{3/2}} \\
            & \ + \sum_{k=0}^\infty \frac{q_2^{(1,k)} r}{[(d_2+h+2kh)^2+r^2]^{3/2}} \Bigg\} \\
            &= \frac{1}{4\pi\epsilon_0} \Bigg\{ \frac{2 q_1}{(1+\epsilon)} \frac{r}{[d_1^2+r^2]^{3/2}} + \sum_{k=0}^\infty \frac{q_1^{(1,k)} r}{[(d_1+2h+2kh)^2+r^2]^{3/2}} \\
            & \ + \sum_{k=0}^\infty \frac{q_2^{(1,k)} r}{[(d_2+h+2kh)^2+r^2]^{3/2}} \Bigg\} .
        \end{split}
    \end{equation}
    \item On the interface, inside Region 0: \\
    \begin{equation}
        \begin{split}
            E_{\parallel 0} &= \frac{q_1^{(0,0)} r}{4\pi\epsilon_0 [d_1^2+r^2]^{3/2}} + \frac{1}{4\pi\epsilon_0} \sum_{k=0}^\infty \Bigg\{ \frac{q_1^{(0,2k+1)} r}{[(d_1+2h+2kh)^2+r^2]^{3/2}} + \frac{q_1^{(0,2k+2)} r}{[(d_1+2h+2kh)^2+r^2]^{3/2}} \Bigg\} \\
            & \ + \frac{1}{4\pi\epsilon_0}  \sum_{k=0}^\infty \Bigg\{ \frac{q_2^{(0,2k)} r}{[(d_2+h+2kh)^2+r^2]^{3/2}} + \frac{q_2^{(0,2k+1)} r}{[(d_2+h+2kh)^2+r^2]^{3/2}} \Bigg\} \\
            &= \frac{1}{4\pi\epsilon_0} \Bigg\{ q_1^{(0,0)} \frac{r}{[d_1^2+r^2]^{3/2}} + \sum_{k=0}^\infty \frac{ \big[ q_1^{(0,2k+1)} + q_1^{(0,2k+2)} \big] r}{[(d_1+2h+2kh)^2+r^2]^{3/2}} \\
            & \ + \sum_{k=0}^\infty \frac{ \big[ q_2^{(0,2k)} + q_2^{(0,2k+1)} \big] r }{[(d_2+h+2kh)^2+r^2]^{3/2}} \Bigg\} .
        \end{split}
    \end{equation}
\end{itemize}

Since $\frac{2q_1}{1+\epsilon} = q_1^{(0,0)}$, $q_1^{(1,k)} = q_1^{(0,2k+1)} + q_1^{(0,2k+2)}$ and $q_2^{(1,k)} = q_2^{(0,2k)} + q_2^{(0,2k+1)}$, we get $E_{\parallel 1} = E_{\parallel 0}$.

\subsection{Normal Condition: $E_{\perp 1} = \epsilon E_{\perp 0}$} 

\begin{itemize}
    \item On the interface, inside Region 1: \\
    \begin{equation}
        \begin{split}
            E_{\perp 1} &= \frac{1}{4\pi\epsilon_0} \Bigg\{ - \frac{q_1 d_1}{[d_1^2+r^2]^{3/2}} + \frac{q_1' d_1}{[d_1^2+r^2]^{3/2}} + \sum_{k=0}^\infty  \frac{q_1^{(1,k)} (d_1+2h+2kh)}{[(d_1+2h+2kh)^2+r^2]^{3/2}} \\
            & \ + \sum_{k=0}^\infty \frac{q_2^{(1,k)} (d_2+h+2kh)}{[(d_2+h+2kh)^2+r^2]^{3/2}} \Bigg\} \\
            &= \frac{1}{4\pi\epsilon_0} \Bigg\{ - \frac{2\epsilon}{1+\epsilon} \frac{q_1 d_1}{[d_1^2+r^2]^{3/2}} + \sum_{k=0}^\infty \frac{q_1^{(1,k)} (d_1+2h+2kh)}{[(d_1+2h+2kh)^2+r^2]^{3/2}} \\
            & \ + \sum_{k=0}^\infty \frac{q_2^{(1,k)} (d_2+h+2kh)}{[(d_2+h+2kh)^2+r^2]^{3/2}} \Bigg\} .
        \end{split}
        \label{E_perp1}
    \end{equation}
\item On the interface, inside Region 0: \\
    \begin{equation}
        \begin{split}
            E_{\perp 0} &= - \frac{q_1^{(0,0)} d_1}{4\pi\epsilon_0 [d_1^2+r^2]^{3/2}} + \frac{1}{4\pi\epsilon_0} \sum_{k=0}^\infty \Bigg\{ \frac{q_1^{(0,2k+1)} (d_1+2h+2kh) }{[(d_1+2h+2kh)^2+r^2]^{3/2}} - \frac{q_1^{(0,2k+2)} (d_1+2h+2kh) }{[(d_1+2h+2kh)^2+r^2]^{3/2}} \Bigg\} \\
            & \ + \frac{1}{4\pi\epsilon_0}  \sum_{k=0}^\infty \Bigg\{ \frac{q_2^{(0,2k)} (d_2+h+2kh)}{[(d_2+h+2kh)^2+r^2]^{3/2}} - \frac{q_2^{(0,2k+1)} (d_2+h+2kh)}{[(d_2+h+2kh)^2+r^2]^{3/2}} \Bigg\} \\
            &= \frac{1}{4\pi\epsilon_0} \Bigg\{ -q_1^{(0,0)} \frac{d_1}{[d_1^2+r^2]^{3/2}} + \sum_{k=0}^\infty \frac{ \big[ q_1^{(0,2k+1)} - q_1^{(0,2k+2)} \big] (d_1+2h+2kh)}{[(d_1+2h+2kh)^2+r^2]^{3/2}} \\
            & \ + \sum_{k=0}^\infty \frac{ \big[ q_2^{(0,2k)} - q_2^{(0,2k+1)} \big] (d_2+h+2kh) }{[(d_2+h+2kh)^2+r^2]^{3/2}} \Bigg\} .
        \end{split}
    \end{equation}
\end{itemize}

Since $\frac{2 \epsilon q_1}{1+\epsilon} = \epsilon q_1^{(0,0)}$, $q_1^{(1,k)} = \epsilon \big[ q_1^{(0,2k+1)} - q_1^{(0,2k+2)} \big]$ and $q_2^{(1,k)} = \epsilon \big[ q_2^{(0,2k)} - q_2^{(0,2k+1)} \big]$, we get $E_{\perp 1} = \epsilon E_{\perp 0}$. 

\section{Confirmation of Newton's third law}

We need to take into account the total forces acting on the dielectric slab, which is equal to $f_1' - f_2'$ where $f_1'$ is the electrostatic force of point-charge $q_1$ exerts on the slab pulling toward and $f_2'$ is the electrostatic force of point-charge $q_2'$ exerts on the slab pulling toward. From Newton's third law, all internal forces need to cancel out, which means we need to prove that:
\begin{equation}
f_1 + (f_1' - f_2') - f_2 = 0 \ .
\label{cancel_out}
\end{equation}

The surface charge density $\sigma_1$ on the interface between Region 1 and Region 0 are given by:
\begin{equation}
\sigma_1 = -\epsilon \left( E_{\perp 1} - E_{\perp 0} \right) = -\epsilon_0 \left( 1 - \frac1{\epsilon} \right) E_{\perp 1} \ ,
\end{equation}
in which $E_{\perp 1}$ can be calculated as described in Eq. \eqref{E_perp1}. Similarly, we can obtain $\sigma_2$.

The force exerted by $q_1$ pulling the dielectric slab (acting on the surface charges on both interfaces) toward it is given by:
\begin{equation}
    \begin{split}
        f_1' &= -\int_0^\infty \frac{q_1}{4 \pi \epsilon_0} \cdot \frac{ 2 \pi \sigma_1 r dr}{d_1^2+r^2} \cdot \frac{d_1}{\sqrt{d_1^2+r^2}} - \int_0^\infty \frac{q_1}{4 \pi \epsilon_0} \cdot \frac{ 2 \pi \sigma_2 r dr}{(d_1+h)^2+r^2} \cdot \frac{d_1+h}{\sqrt{(d_1+h)^2+r^2}} \\
        &= \frac{q_1}{4 \pi \epsilon_0} \int_0^\infty \frac{1}{2} \left( 1- \frac{1}{\epsilon} \right) \Bigg\{ - \frac{2\epsilon}{1+\epsilon} \frac{q_1 d_1}{[d_1^2+r^2]^{3/2}} + \sum_{k=0}^\infty \frac{q_1^{(1,k)} (d_1+2h+2kh)}{[(d_1+2h+2kh)^2+r^2]^{3/2}} \\
        & \ + \sum_{k=0}^\infty \frac{q_2^{(1,k)} (d_2+h+2kh)}{[(d_2+h+2kh)^2+r^2]^{3/2}} \Bigg\} \frac{d_1 rdr}{(d_1^2+r^2)^{3/2}} + \frac{q_1}{4 \pi \epsilon_0} \int_0^\infty \frac{1}{2}  \left( 1- \frac{1}{\epsilon} \right) \Bigg\{ - \frac{2\epsilon}{1+\epsilon} \frac{q_2 d_1}{[d_2^2+r^2]^{3/2}} \\
        & \ + \sum_{k=0}^\infty \frac{q_2^{(2,k)} (d_2+2h+2kh)}{[(d_2+2h+2kh)^2+r^2]^{3/2}} + \sum_{k=0}^\infty \frac{q_1^{(2,k)} (d_1+h+2kh)}{[(d_1+h+2kh)^2+r^2]^{3/2}} \Bigg\} \frac{(d_1+h)rdr}{[(d_1+h)^2+r^2]^{3/2}} .
    \end{split}
\end{equation}
Here, we note an useful integration:
\begin{equation}
 \int_0^\infty \frac{abxdx}{(a^2+x^2)^{3/2} (b^2+x^2)^{3/2} } = \frac{1}{(a+b)^2} \ , 
\end{equation}
where $a$ and $b$ are positive real numbers. Hence, with that, we obtain:
\begin{equation}
    \begin{split}
        f_1' &= \frac{q_1}{4 \pi \epsilon_0} \frac{1}{2} \left( 1- \frac{1}{\epsilon} \right) \Bigg\{ - \frac{2\epsilon}{1+\epsilon} \frac{q_1}{(2d_1)^2} + \sum_{k=0}^\infty \frac{q_1^{(1,k)}}{(2d_1+2h+2kh)^2} + \sum_{k=0}^\infty \frac{q_2^{(1,k)}}{(d_1+d_2+h+2kh)^2} \Bigg\} \\
        & \ + \frac{q_1}{4 \pi \epsilon_0} \frac{1}{2} \left( 1- \frac{1}{\epsilon} \right) \Bigg\{ - \frac{2\epsilon}{1+\epsilon} \frac{q_2}{(d_1+d_2+h)^2} + \sum_{k=0}^\infty \frac{q_2^{(2,k)}}{(d_1+d_2+h+2h+2kh)^2} + \sum_{k=0}^\infty \frac{q_1^{(2,k)}}{(2d_1+2h+2kh)^2} \Bigg\} \\
        &= \frac{q_1}{4 \pi \epsilon_0} \Bigg\{ - \frac{\epsilon-1}{\epsilon+1} \frac{q_1}{(2d_1)^2} + \sum_{k=0}^\infty \frac{1}{2} \left( 1- \frac{1}{\epsilon} \right) \frac{q_1^{(1,k)}+q_1^{(2,k)}}{(2d_1+2h+2kh)^2} + \sum_{k=0}^\infty \frac{1}{2} \left( 1- \frac{1}{\epsilon} \right) \frac{q_2^{(1,k+1)} +q_2^{(2,k)}}{(d_1+d_2+h+2h+2kh)^2} \\
        & \ + \left( - \frac{\epsilon-1}{\epsilon+1} q_2 + \frac{1}{2} \left( 1- \frac{1}{\epsilon} \right) q_2^{(1,0)} \right) \frac{1}{(d_1+d_2+h)^2} \Bigg\}.
    \end{split}
\end{equation}

Since $ \frac{1}{2} \left( 1- \frac{1}{\epsilon} \right) \left( q_1^{(1,k)}+q_1^{(2,k)} \right) = q_1^{(1,k)} $, $ \frac{1}{2} \left( 1- \frac{1}{\epsilon} \right) \left( q_2^{(1,k+1)}+q_1^{(2,k)} \right) = q_2^{(1,k+1)} $ and $ - \frac{\epsilon-1}{\epsilon+1} q_2 + \frac{1}{2} \left( 1- \frac{1}{\epsilon} \right) q_2^{(1,0)} = q_2^{(1,0)} -q_2 $, from Eq. (7) in the main manuscript we get:
\begin{equation}
    f_1 - f_1' = \frac{q_1 q_2}{4 \pi \epsilon_0 (d_1 + h + d_2)^2} \ .
\end{equation}
Similarly we can obtain $f_2 - f_2'$, which is the negative of the above expression, thus verify Eq. \eqref{cancel_out}.

\section{Solving the Poisson's Equation}

Consider the simple case where the setting is symmetric, $q_1=q_2=q$ and $d_1=d_2=d$. Choose a Cartesian $Oxyz$ coordinate system, in which the origin is inside the dielectric slab at the middle of two charges, the $xy$-plane is parallel to the slab thus the $z$-axis passes through both charges. Decompose the potential into Fourier-modes in the $xy$-plane:
\begin{equation}
V(x,y,z) = \int \frac{dk_x dk_y}{(2\pi)^2} e^{ik_x x + ik_y y} \tilde{V}_{k}(z) \ ,
\end{equation}
where $k=\sqrt{k_x^2 + k_y^2}$ is a sufficient index due to rotational symmetry in the $xy$-plane. Excluding the two interfaces and the positions of two charges we get the Laplace's equation $\nabla^2 V = 0$, hence the solution will be of the form (due to the setting is symmetric and far-away potential should go to 0):
\begin{equation}
\begin{split}
z > d + \frac{h}2 \ &: \ \tilde{V}_k(z) = A e^{-kz} \ ,
\\
d + \frac{h}2 > z > \frac{h}2 \ &: \ \tilde{V}_k(z) = B e^{-kz} + C e^{+kz} \ ,  
\\
\frac{h}2 > |z| \ &: \ \tilde{V}_k(z) = D \left( e^{-kz} + e^{+kz} \right) \ ,
\\
d+\frac{h}2 > -z >\frac{h}2  \ &: \ \tilde{V}_k(z)  = C e^{-kz} + B e^{+kz} \ , 
\\
-z>d+\frac{h}2 \ &: \ \tilde{V}_k(z)  = A e^{+kz} + D e^{+kz} \ .
\end{split}
\end{equation}
The continuity conditions of $\tilde{V}_k(z)$ at the charges' positions and the  interfaces give:
\begin{equation}
\begin{split}
& A e^{-k(d+h/2)} = B e^{-k(d+h/2)} + C e^{-k(d+h)/2} \ ,
\\
&B e^{-kh/2} + C e^{+kh/2} = D \left(e^{+kh/2} + e^{+kh/2} \right) \ .
\end{split}
\label{cont_value}
\end{equation}
The jumping conditions of $\partial_z \tilde{V}_k(z)$ at the charges' positions give:
\begin{equation}
-\frac{q}{\epsilon_0} = \left( - A k e^{-k(d+h/2)} \right) - \left( - B k e^{-k(d+h/2)} + C k e^{+k(d+h/2)} \right) \ .
\label{cont_grad_charge}
\end{equation}
The continuity conditions of $\partial_z \tilde{V}_k(z)$ at the interfaces give:
\begin{equation}
0 = \left( - B k e^{-kh/2} + C k e^{+kh/2} \right) - \epsilon D \left( - k e^{-kh/2} + k e^{+kh/2}\right) \ .
\label{cont_grad_interface}
\end{equation}
Four unknowns $A$, $B$, $C$, $D$ can be solved with four equations \eqref{cont_value}, \eqref{cont_grad_charge}, \eqref{cont_grad_interface}: 
\begin{equation}
\begin{split}
A & = \frac{2 q e^{kh} \left(\cosh(kd)\cosh(\frac{kh}2) + \epsilon \sinh(kd)\sinh(\frac{kh}2) \right)}{\epsilon_0 k\left( (1 - \epsilon) + (1+\epsilon) e^{kh} \right)}\ , 
\\
B & = \frac{q e^{-k(d - h/2)} \left( (1 + \epsilon) + (1 - \epsilon) e^{kh} \right)}{2\epsilon_0 k\left( (1 - \epsilon) + (1+\epsilon) e^{kh} \right)} \ , 
\\
C & = \frac{q e^{-k(d+h/2)}}{2\epsilon_0 k} \ , \\ 
D  & = \frac{q e^{-k(d - h/2)}}{\epsilon_0 k \left( (1 - \epsilon) + (1+\epsilon) e^{kh} \right)} \ .
\end{split}
\end{equation}
Now let's look at the charge's position $z=d+h/2$ and the self-contribution:
\begin{equation}
\begin{split}
z > d + \frac{h}2 \ &: \ \tilde{V}^{(s)}_k(z) = A^{(s)}_+ e^{-kz} \ ,
\\
d + \frac{h}2 > z \ &: \ \tilde{V}^{(s)}_k(z)  = A^{(s)}_- e^{+kz} \ .
\end{split}
\end{equation}
The continuity condition of $\tilde{V}^{(s)}_k(z)$:
\begin{equation}
A^{(s)}_+ e^{-k(d+h/2)} = A^{(s)}_- e^{+k(d+h/2)} \ . 
\label{cont_value_self}
\end{equation}
The jumping conditions of $\tilde{V}^{(s)}_k(z)$:
\begin{equation}
-\frac{q}{\epsilon_0} = \left( - A^{(s)}_+ k e^{-k(d+h/2)} \right) - \left(  A^{(s)}_- k e^{+k(d+h/2)} \right) \ .
\label{cont_grad_self}
\end{equation}
From \eqref{cont_value_self} and \eqref{cont_grad_self}, we get:
\begin{equation}
A^{(s)}_{\pm} = \frac{q e^{\pm k(d+h/2)}}{2\epsilon_0 k}
\end{equation}
The regularized $\tilde{V}^{(r)}_k(z) = \tilde{V}_k(z) - \tilde{V}^{(s)}_k(z)$ is continuous and smooth at that charge's position, and can be used to determine the gradient right there:
\begin{equation}
\begin{split}
\partial_z \tilde{V}^{(r)}_k(z) & \Big|_{z = d + \frac{h}2} = \left( A - A^{(s)}_+ \right) \partial_z e^{-kz} \Big|_{z = d + \frac{h}2}
\\
\ \ \ \ \  & = - \frac{q e^{-2kd}}{2\epsilon_0} \frac{1 - \epsilon \tanh(\frac{kh}2)}{1 + \epsilon \tanh(\frac{kh}2)} \ .
\end{split}
\end{equation}
Thus the force acting on the charge can be calculated with:
\begin{equation}
\begin{split}
f & = - q \partial_z V^{(r)}(0,0,z) \Big|_{z = d+\frac{h}2} 
\\
& = - q \int^{\infty}_0 \frac{2\pi k dk}{(2\pi)^2} \partial_z \tilde{V}^{(r)}_k(z) \Big|_{z = d+\frac{h}2}
\\
& = \frac{q^2}{4\pi \epsilon_0} \int_0^{\infty} dk k e^{-2kd} \frac{1 - \epsilon \tanh(\frac{kh}2)}{1 + \epsilon \tanh(\frac{kh}2)} \ .
\end{split}
\end{equation}
Define $\chi=kh/2$ then:
\begin{equation}
f = \frac{q^2}{4\pi \epsilon_0 h^2} \times 4 \int_0^{\infty} d\chi \chi e^{-4\chi d/h} \frac{1 - \epsilon \tanh \chi }{1 + \epsilon \tanh \chi} \ .
\end{equation}
When $d=h$, we get back the result Eq. (13) in the main manuscript.

\end{document}